\documentclass[pre,10pt,twocolumn,showpacs]{revtex4}  
\usepackage[dvips]{epsfig}
\usepackage{float}
\usepackage{amsmath}
\usepackage{amssymb}
\usepackage{pslatex}
\usepackage{hhline}
\usepackage{threeparttable}

\newcommand{\vrr}{\mathbf{r}}

\newcommand{\vecb}[1]{\mathbf{#1}}

\begin{document}
\title{Nonlinear absorption and dispersion in fiber-taper-coupled silicon photonic crystal microresonators}
\author{Paul E. Barclay}
\email{pbarclay@caltech.edu}
\author{Kartik Srinivasan}
\author{Oskar Painter}
\affiliation{Department of Applied Physics, California Institute of Technology, Pasadena, CA 91125, USA.}
\date{\today}

\begin{abstract}
A technique is demonstrated which efficiently transfers light between a tapered standard single-mode optical fiber and a high-Q, ultra-small mode volume, silicon photonic crystal resonant cavity.   Cavity mode quality factors of $4.7\times 10^4$ are measured, and a total fiber-to-cavity coupling efficiency of 44\% is demonstrated.  Using this efficient cavity input and output channel, the steady-state nonlinear absorption and dispersion of the photonic crystal cavity is studied.  Optical bistability is observed for fiber input powers as low as $250$ $\mu$W, corresponding to a dropped power of $100$ $\mu$W and $3$ fJ of stored cavity energy.  A high-density effective free-carrier lifetime for these silicon photonic crystal resonators of $\sim 0.5$ ns is also estimated from power dependent loss and dispersion measurements.
\end{abstract}

\maketitle

\section{Introduction}
\label{sec:Intro}
Recently it has been demonstrated that resonant microcavities formed in planar photonic crystals (PC) are capable of not only confining light to ultra-small optical mode volumes, but also of realizing photon cavity lifetimes large enough \cite{ref:Srinivasan3,ref:Noda4,ref:Srinivasan4} to enable, for instance, reaching strong-coupling with atomic Cs \cite{ref:Lev} or semiconductor quantum dots \cite{ref:Yoshie2,ref:Reithmaier,ref:Peter}.  The enhancement to the local energy density enabled by PC cavities is also of significant interest in nonlinear optics, as it reduces the input power required for nonlinear effects, such as optical bistability \cite{ref:Cowan3,ref:Soljacic2}.  However, in addition to requiring large field enhancements and photon lifetimes, many of the proposed applications of PC cavities in quantum, nonlinear, and integrated optics, demand the ability to efficiently interface the PC cavity with external optics.  For example, using high-Q PC cavities for chip based cavity-QED (cQED) \cite{ref:Lev} or in single-photon sources \cite{ref:Gerard3}, where photon collection is an important measure of device performance \cite{ref:Brassard1,ref:Knill,ref:Kok}, requires an efficient coupling scheme to the sub-micron cavity mode. Similarly, proposed applications of microresonators in nonlinear-optical switching circuits \cite{ref:Soljacic3} puts a premium on reduced input power and efficient optical coupling techniques.  

The difficulty in optically accessing PC cavities is largely a result of their ultra-small mode volume and external radiation pattern, which unlike micropost \cite{ref:Pelton} and Fabry-P\'{e}rot \cite{ref:McKeever1} cavities, is not inherently suited to coupling with conventional free-space or fiber optics.  Only recently have low-loss fiber coupling techniques to planar photonic crystal waveguides been demonstrated experimentally, using on-chip spot size converters ($\sim$ 1-4 dB loss/port \cite{ref:McNab,ref:Notomi3}), out-of plane diffraction gratings ($\sim$ 8-10 dB loss/port \cite{ref:Baets2,ref:Cowan2}), and narrow-band evanescent coupling ($\sim$ 0.1 dB loss/port \cite{ref:Barclay5}).  In this paper we employ the latter technique to efficiently source and collect light from a silicon (Si) high-Q PC cavity via a photonic crystal waveguide (PCWG) \cite{ref:Loncar} which is evanescently coupled to a fiber taper \cite{ref:Knight}.  Using a mode-matched PCWG-PC cavity design \cite{ref:Barclay2} to minimize parasitic loading of the PC cavity, cavities loaded to $60\%$ of critical and maintaining a $Q$-factor close to $4 \times 10^4$ are demonstrated, and a total fiber-to-cavity coupling efficiency of $44\%$ is measured.  The utility of this efficient fiber coupling technique is then demonstrated in studies of the power dependent nonlinear response of the PC cavity. We observe optical bistability for $100$ $\mu\text{W}$ dropped cavity power, and predict sub-nanosecond free carrier lifetimes in the cavity.

An outline of the paper is as follows.  We begin in Section \ref{sec:couple_theory_linear} with a description of the fiber-to-cavity coupling scheme, with specific emphasis on issues associated with the complex modal properties of a photonic crystal microcavity.  In Section \ref{sec:NL_theory} we augment this linear theory by incorporating nonlinear processes which depend upon the magnitude of the internally stored cavity energy.  Low power measurements of a fabricated silicon photonic crystal waveguide-cavity system are presented in Section \ref{sec:Lin_meas}, where optical fiber taper probing is used to both confirm the localized nature of a PC cavity resonance and to study the losses in the fiber-cavity system.  Higher power measurements in which  nonlinear effects become apparent are studied in Section \ref{sec:NL_meas}, and the model presented in Section \ref{sec:NL_theory} is used to estimate the scale of the different nonlinear processes within the silicon PC cavity.  Finally, a summary is given in Section \ref{sec:Summary}.
     
\begin{figure}[ht]
\begin{center}
\epsfig{figure=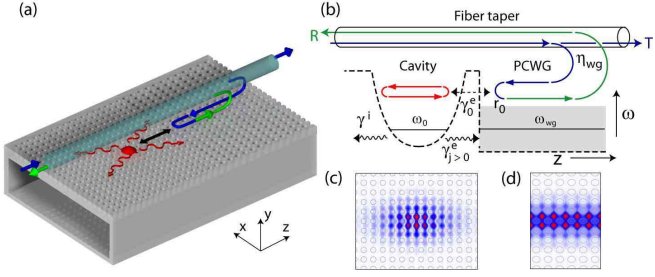, width=1.0\linewidth}
\caption{(a) Schematic of the fiber taper to PC cavity coupling scheme.  The blue arrow represents the input light, some of which is coupled contradirectionally into the PCWG. The green arrow represents the light reflected by the PC cavity and recollected in the backwards propagating fiber mode.  The red colored region represents the cavity mode and its radiation pattern. (b) Illustration of the fiber-PC cavity coupling process.   The dashed line represents the ``local'' band-edge frequency of the photonic crystal along the waveguide axis.   The step discontinuity in the bandedge at the PCWG - PC cavity interface is due to a jump in the longitudinal ($\hat{z}$) lattice constant (see Fig.\ \ref{fig:SEM}).   The parabolic ``potential''  is a result of the longitudinal grade in hole radius of the PC cavity.  The bandwidth of the waveguide is represented by the gray shaded area.  Coupling between the cavity mode of interest (frequency $\omega_0$) and the mode matched PCWG mode ($\omega_{\text{WG}} = \omega_0$) is represented by $\gamma^e_0$, coupling to radiating PCWG modes is represented by $\gamma^e_{j>0}$, and intrinsic cavity loss is represented by $\gamma^i$.  (c,d) Magnetic field profile, calculated using FDTD, of the high-Q PC cavity $A_{2}^0$ mode and the fundamental $\text{TE}_{1}$ PCWG mode, respectively.
}
\label{fig:illustration}
\end{center}
\end{figure} 

\section{Coupling scheme - theory}
\label{sec:couple_theory_linear}

An illustration of the coupling scheme is shown in Figs. \ref{fig:illustration}(a) and \ref{fig:illustration}(b).  In this scheme, evanescent coupling between an optical fiber taper and a PCWG  is used to interface with the PC chip.  Once on the chip, light is guided to a PC microcavity at the terminus of the PCWG.  Light that is reflected from the PC cavity is then recollected into the backward propagating fiber taper mode, where it is separated from the forward propagating input signal using a fiber splitter.  In this way a single optical fiber is used to both source and collect light from the PC microcavity.

The PC waveguide and PC microcavity studied here are formed from a two-dimensional photonic crystal consisting of a square lattice array of air holes in an optically thin slab waveguide.  This PC cavity-waveguide system was previously studied theoretically in Ref. \cite{ref:Barclay2}, where the fundamental ($\text{TE}_{1}$) mode of the PCWG was designed to mode-match with the fundamental ($A_{2}^0$) cavity mode.  The mode pattern of the TE$_{1}$ PCWG mode and the $A_{2}^0$ cavity mode, calculated using the finite-difference time-domain (FDTD) method, are shown in Fig. \ref{fig:illustration}(c) and Fig. \ref{fig:illustration}(d), respectively.  The mode-matched cavity acts as a mirror with high modal reflectivity, $r_o(\omega)$, except at frequencies of the localized  cavity states, where light can resonantly tunnel between the PCWG and the cavity.  Previous studies have shown that the optical fiber-PCWG evanescent coupler has near unity coupling efficiency ($97\%$) over a bandwidth of roughly 15 nm \cite{ref:Barclay5}, thus providing the necessary efficient fiber-chip optical interface.  Also, in an independent study, the fundamental $A_{2}^0$ mode of the graded square lattice PC cavity was measured to have a $Q$-factor of $4 \times 10^4$ and a mode localization consistent with an effective mode volume of $V_{\text{eff}} = 0.9(\lambda/n)^3$\cite{ref:Srinivasan4}. 

Here we combine these two complementary components, and study the efficiency with which the PC cavity is loaded by the mode-matched, fiber-coupled PCWG.  We begin with a review of some of the key parameters describing the loading of a general resonant structure, pointing out specific implications for planar PC microcavities. 

\subsection{Efficient waveguide to cavity loading}\label{sec:eff_coupling}

As proposed in Ref. \cite{ref:Spillane2} in the context of microsphere resonators, the interaction between a PC cavity and an external PC waveguide can be described by two key parameters, the \emph{coupling parameter} $K$ and the \emph{ideality factor} $I$:
\begin{align}
K &\equiv \frac{\gamma^{e}_{0}}{\gamma^{i} + \sum_{j \ne 0} \gamma^{e}_{j}}, \label{eq:coupling_factor}\\
I &\equiv \frac{\gamma^{e}_{0}}{\sum_{j} \gamma^{e}_{j}},\label{eq:ideality_factor}
\end{align}
\noindent where the cavity mode is characterized by its resonance frequency $\omega_o$, its intrinsic photon loss rate in absence of the external PCWG ($\gamma^{i}$), and its coupling rates to the fundamental $\text{TE}_{1}$ mode ($\gamma^{e}_{0}$) and higher order (including radiating) modes of the external PCWG ($\gamma^{e}_{j>0}$).  $I$ describes the degree of ``good'' loading, via the PCWG $\text{TE}_1$ mode in this case, relative to the total loading of the resonator.  $K$, on the other hand, is the ratio of ``good'' loading to the parasitic and intrinsic loss channels of the resonator. 

The coupling parameter $K$ determines the on-resonance fraction of optical power reflected by the cavity back into the PCWG mode, 
\begin{equation}
R_{o}(\omega_{o}) = \frac{(1-K)^2}{(1+K)^2}.
\end{equation}
The remaining fractional power, $1-R_{o}(\omega_{o})$, is absorbed inside the PC cavity or radiated into the parasitic output channels.  The reflection resonance full-width at half-maximum (FWHM) linewidth is given by the sum of the loss rates for \emph{all} of the loss channels of the cavity, $\delta\omega = \gamma^i + \sum_{j} \gamma^{e}_{j}$.  From $R_{o}(\omega_{o})$ and $\delta\omega$ the quality factor of the PC cavity mode due to intrinsic and parasitic loss (i.e., those loss channels other than the ``good'' PCWG $\text{TE}_{1}$ channel) can be determined,
\begin{equation}\label{eq:Q_i+P}
Q_{i+P} = 2Q_{T}\frac{1}{1 \pm \sqrt{R_{o}(\omega_{o})}} = Q_{T}(1+K),
\end{equation}
\noindent where the total loaded quality factor is $Q_{T} = \omega_{o}/\delta\omega$, and where the $\pm$ corresponds to the under- and over- coupled ($K \lessgtr 1$) loading condition.  On resonance, full power transfer (critical coupling) from the ``good'' loading channel to the resonant PC cavity mode occurs when $K=1$.  

Whereas $K$ determines the amount of power dropped by the resonator, the role of $I$ is more subtle.  In the case of an internal emitter, the collection efficiency ($\eta_{0}$) of emitted photons into the ``good'' loading channel is given by,
\begin{equation}\label{eq:eta_0}
\eta_{0} = \frac{\gamma^e_{0}}{\gamma^i + \sum_{j} \gamma^{e}_{j}} = \frac{1}{1+1/K}.
\end{equation}
\noindent which depends only upon the coupling parameter $K$.  However, the cost of obtaining a large collection efficiency is measured by the drop in loaded quality factor of the resonant cavity mode, which can be written in terms of $K$, $I$, and $Q_{i}$ as 
\begin{equation}\label{eq:Q_T}
\frac{Q_{T}}{Q_{i}} = 1 - \frac{K}{(I(1+K))} = 1 - \frac{\eta_{0}}{I}.  
\end{equation}
\noindent Thus, for a given collection efficiency, to maintain a long cavity photon lifetime, $I$ should be maximized.  

Utilizing a cavity loading method with $I \sim 1$ is also important for cavity based nonlinear optics.  A simple argument can be made by studying the stored energy inside a resonant cavity for a given input power.  One can write for the on-resonance internal stored energy $U$,
\begin{equation}\label{eq:U}
U = (1 - R_o(\omega_o)) \frac{Q_{i+P}}{\omega_{o}}  P_{i} = \frac{4K}{(1+K)^2}\frac{I-K(1-I)}{I}\frac{Q_i}{\omega_o}P_i
\end{equation}
\noindent where $P_{i}$ is the input power in the ``good'' loading channel.  The maximum stored energy in the resonator occurs at $K_{\text{max}} = I/(2-I)$, giving a peak stored energy $U_{\text{max}} = I(Q_{i}/\omega_{o})P_{i}$, which scales directly with $I$.  

The integrated PC cavity-waveguide design employed here has two important features which serve to maximize $I$:  (i) the waveguide and cavity modes have similar transverse field profiles (see Figs. \ref{fig:illustration}(c-d)) which allows the cavity to be efficiently loaded end-on, and (ii) the end-fire PCWG-cavity geometry restricts the cavity to a single dominant output channel, in contrast to side-coupled \cite{ref:Noda4}, in-line \cite{ref:Lin4}, and direct taper coupled \cite{ref:Srinivasan4} geometries, in which the cavity radiates equally into backward and forward propagating waveguide modes (bounding $K \le 1$ and $I \le 0.5$).  In comparison to other microcavity systems, the geometry of the PC cavity-waveguide studied here is analogous to a Fabry-P\'{e}rot cavity with a high reflectivity back mirror and a lower reflectivity front mirror through which a mode-matched input beam sources the cavity.  The geometry is also similar to a side-coupled traveling wave resonator, such as a microsphere or microdisk that supports whispering-gallery type modes, in which the traveling wave resonance radiates selectively into a single phase matched output channel.              

\section{Influence of nonlinear absorption and dispersion on cavity response - theory}
\label{sec:NL_theory}
Owing to the ultra-small mode volume and long resonant photon lifetimes of the PC cavities studied here, the resulting stored electromagnetic energy density can be extremely large even for modest input powers ($< \text{mW}$), resulting in highly nonlinear behavior of the resonant cavity system.  In order to account for nonlinear effects one may modify the (linear) analysis of Section \ref{sec:eff_coupling} by allowing the various cavity and coupling parameters to depend upon the stored cavity energy, a reasonably easy quantity to estimate from experimental measurements.  In this section, relevant nonlinear processes are explicitly incorporated into the description of the cavity response through use of carefully defined effective modal volumes and confinement factors appropriate to nonlinear processes in high-index contrast photonic crystal structures.  We begin with a description of nonlinear absorption, which tends to drive the steady-state nonlinear response of the PC cavities studied here.    

\subsection{Nonlinear absorption}
\label{subsec:NL_abs_theory}
Nonlinear absorption adds power dependent loss channels to the photonic crystal cavity, degrading the quality factor as the internal cavity energy  is increased, which in turn modifies the coupling efficiency from the PCWG loading channel.  This effect is incorporated into the formalism presented in Section \ref{sec:eff_coupling} by writing the intrinsic cavity loss rate, $\gamma^i$, explicitly in terms of its various linear and nonlinear components:
\begin{equation}
\label{eq:general_gamma_i}
\gamma^i(U) = \gamma_{\text{rad}} + \gamma_{\text{lin}} + \overline{\gamma}_{\text{TPA}}(U) + \overline{\gamma}_{\text{FCA}}(U).
\end{equation}
\noindent At low power, the ``cold cavity'' loss rate is given by $\gamma_{\text{rad}}$ and $\gamma_{\text{lin}}$, which represent loss due to radiation and linear material absorption, respectively.  Power dependent nonlinear loss is given here by $\overline{\gamma}_{\text{TPA}}$ and $\overline{\gamma}_{\text{FCA}}$, which represent two-photon and free-carrier absorption, respectively; other nonlinear absorption processes can be included analogously.  The coupling parameter $K$ and  the quality factor $Q_{i+P}$ depend on $\gamma^i$, requiring the solution of a system of self-consistent equations for $U$ in order to determine the on-resonance cavity response for a given PCWG input power $P_i$:
\begin{align}
U &= \frac{4 K(U)}{(1+K(U))^2} \frac{Q_{i+P}(U)}{\omega_{o}}  P_{i}, \label{eq:U_NL}\\
K(U) &= \frac{\gamma_o^e}{\gamma^i(U) + \sum_{j>0} \gamma_j^e}, \label{eq:K_NL}\\
\frac{\omega_o}{Q_{i+P}(U)} &= \gamma^i(U) + \sum_{j>0} \gamma_j^e \label{eq:Q_i_P_NL}.
\end{align}
\noindent Before eqs. (\ref{eq:U_NL} - \ref{eq:Q_i_P_NL}) can be solved, explicit expressions for the energy-dependent contributions to $\gamma^i$ are required.   Beginning with relations for nonlinear absorption in bulk media, and taking into account the complicated geometry of the PC cavity, we now derive expressions for $\overline{\gamma}_{\text{TPA}}$ and $\overline{\gamma}_{\text{FCA}}$.  These expressions can be written in terms of the internal cavity energy, known material parameters, and modal parameters which account for the mode shape and localization of the PC cavity field.

\subsubsection{Two-Photon Absorption}
\label{subsubsec:TPA_theory}
The 1500 nm operating band of the devices studied in this work lies in the bandgap of the host silicon material.  For the doping densities of the p-type silicon membrane used to form the PC cavity ($\rho \sim 1-3$ $\Omega \cdot \text{cm}$, $N_{A} < 10^{16}$  $\text{cm}^{-3}$), free-carrier absorption due to ionized dopants is small ($\alpha_{fc} \sim 10^{-2}$  cm$^{-1}$).  Two-photon absorption however, is significant\cite{ref:Dinu,ref:Liang,ref:Kanamoto,ref:Cowan2}, especially in the highly localized PC cavities.  For a given field distribution, the (time-averaged) two-photon absorption loss rate at position $\vrr$ can be written as
\begin{equation}
\label{eq:TPA_bulk}
\gamma_{\text{TPA}}(\vrr) = \beta'(\vrr) \frac{1}{2} \epsilon_o n^2(\vrr)E^2(\vrr),
\end{equation}
\noindent where $E(\vrr)$ is the \emph{amplitude} of the complex electric field pattern $\vecb{E}(\vrr)$ of the resonant mode of the cavity, $\epsilon_o$ is the permittivity of free space, and $n(\vrr)$ is the local (unperturbed) refractive index.  The real, physical electric field of the resonant cavity mode can be written in terms of the complex mode pattern as $\vecb{E}(\vrr,t) = (\vecb{E}(\vrr)e^{-i\omega_o t} + \vecb{E}^{*}(\vrr)e^{+i\omega_o t})/2$.  The \emph{material} parameter, $\beta'$, describes the strength of the two-photon absorption process, and can be related to the usual two-photon absorption coefficient, $\beta$, which relates intensity to loss per unit length, by $\beta' =  (c/n_g)^2\beta$, where $c$ is the speed of light in vacuum and $n_g$ is the group velocity index associated with the measurement of $\beta$.  Typically, for bulk material measurements where waveguiding is minimal and material dispersion is small, $n_g$ can be taken to be equal to $n$.

In high-index-contrast photonic crystals, $E$, $n$, and $\beta'$ depend strongly on spatial coordinate $\vrr$.  Equation (\ref{eq:TPA_bulk}) describes the local two-photon absorption rate;  the effective modal two-photon absorption rate which characterizes the absorption of the entire cavity mode is given by a weighted average of the local absorption rate\footnote{One can show\cite{ref:Agrawal,ref:Johnson6} that the appropriate weighting is $n^2(\vrr)E^2(\vrr)$, proportional to the local electric field energy density.  Note also that the time averaged electromagnetic energy stored in the cavity, $U$, is $\int (1/2) \epsilon_o n^2(\vrr) E^2(\vrr) d\vrr$ for a harmonic mode in a non-dispersive dielectric.}
\begin{align}
\label{eq:gamma_tpa_bar_1}
\overline{\gamma}_{\text{TPA}} &= \frac{\int{ \gamma_{\text{TPA}}(\vrr) n^2(\vrr)E^2(\vrr)d\vrr}} { \int n^2(\vrr)E^2(\vrr)d\vrr} = \overline{\beta'}\frac{U}{V_{\text{TPA}}}, 
\end{align}
\noindent where $\overline{\beta^{\prime}}$ and $V_{\text{TPA}}$ are defined as,
\begin{align}
\overline{\beta^{\prime}} &= \frac{\int{\beta'(\vrr) n^4(\vrr)E^4(\vrr) d\vrr}}{\int{n^4(\vrr)E^4(\vrr)}d\vrr} \label{eq:beta_bar}\\ 
V_{\text{TPA}} &=  \frac{ \left( \int{n^2(\vrr) E^2(\vrr) d\vrr}\right)^2}{\int{n^4(\vrr)E^4(\vrr) d\vrr}} \label{eq:V_TPA}.
\end{align}
\noindent  In a photonic crystal formed by air holes in silicon, $\beta'(\vrr) = \beta'_{\text{Si}}$ inside the silicon and $\beta'(\vrr) = 0$ in the air, so that eq. (\ref{eq:gamma_tpa_bar_1}) can be written as,
\begin{align}
\overline{\gamma}_{\text{TPA}} &= \Gamma_{\text{TPA}}\beta'_{\text{Si}}\frac{U}{V_{\text{TPA}}} \label{eq:gamma_tpa_bar_2}\\
\Gamma_{\text{TPA}} &= \frac{\int_{\text{Si}} n^4(\vrr)E^4(\vrr) d\vrr} {\int{n^4(\vrr)E^4(\vrr)}d\vrr} \label{eq:Gamma_TPA},
\end{align}
\noindent for which  $\int_{\text{Si}}$ only integrates over the silicon region of the PC cavity. 
\subsubsection{Free-carrier absorption}
\label{subsubsec:FCA_theory}
Although, as mentioned above, the (linear) free-carrier absorption due to the ionized dopants of the silicon layer used for the PC cavities in this work is negligible on the scale of other losses, two-photon absorption gives rise to a steady-state population of electron and hole free-carriers far above this equilibrium value. Two-photon absorption induced free-carrier absorption thus plays a significant role in the silicon PC cavity nonlinear response. At position $\vrr$ in the cavity, assuming a simple Drude model, the optical loss rate due to free-carrier absorption is
\begin{equation}
\label{eq:gamma_fca}
\gamma_{\text{FCA}} = \sigma'(\vrr)N(\vrr),
\end{equation}
\noindent where $\sigma'$ is related to the material dependent free-carrier cross-section, $\sigma$, by $\sigma' = \sigma(c/n_{g})$, and $N(\vrr)$ is the free-carrier density.  In silicon it has been demonstrated experimentally\cite{ref:Soref} that this model correctly describes absorption by both electrons and holes, albeit with unique values of $\sigma'_{e,h}$ for each carrier type. Here we let $N$ represent the number of electron-hole pairs\footnote{We neglect the small ($<1\times 10^{16}$ cm$^{-3}$) background free-carrier hole density due to the ionized acceptors of the p-type Si layer used in this work.}, and take $\sigma' = \sigma'_e + \sigma'_h$.  

In general, the derivation of the free-carrier density for a given two-photon absorbed power distribution requires a microscopic theory taking into account carrier diffusion, carrier-carrier scattering effects (Auger recombination for instance), and in the highly porous PC cavities, local surface recombination effects.  In lieu of such an analysis, we approximate the free-carrier density distribution by considering the local two-photon absorbed power,
\begin{equation}
\label{eq:N_fc}
N(\vrr) = \frac{\tau p_{\text{TPA}}(\vrr)}{2\hbar\omega_o},
\end{equation}
\noindent where $\tau$ is a free-carrier lifetime, and $p_{\text{TPA}}(\vrr)$ is the local absorbed power \emph{density} due to two-photon absorption,
\begin{equation}
\label{eq:p_tpa_simple}
p_{\text{TPA}}(\vrr) = \frac{1}{2}\epsilon_o n^2(\vrr) E^2(\vrr)\gamma_{\text{TPA}}(\vrr).
\end{equation}
\noindent Equation (\ref{eq:N_fc}) neglects non-local effects due to spatial carrier diffusion by assuming that $N(\vrr)$ depends only on the power absorbed at position $\vrr$; however, it does correlate regions of strong two-photon absorption with high free-carrier density.  Also, since $\tau$ generally depends on $N$ and on the proximity to surfaces, $\tau$ will have a spatial dependence within the cavity.  We neglect this effect here, and let $\tau$ represent an effective free-carrier lifetime for all the carriers in the cavity region\footnote{In using the approximate theory above, in which regions of high two-photon absorbed power are correlated with high steady-state carrier density, we better approximate the cavity ``volume'' of interest, and consequently the effective free-carrier lifetime better represents the average time a free-carrier stays in the region of the PC cavity mode.}.  Combining eqs. (\ref{eq:gamma_fca}), (\ref{eq:N_fc}) and (\ref{eq:p_tpa_simple}), an effective modal free-carrier absorption rate can be written as
\begin{align}
\overline{\gamma}_{\text{FCA}} &=  \frac{\frac{\tau}{2\hbar\omega_o} \int{ \left(\sigma'(\vrr)\frac{1}{2} \epsilon_o n^2(\vrr)E^2(\vrr) \gamma_{\text{TPA}}(\vrr) \right) n^2(\vrr)E^2(\vrr)d\vrr}}{\int n^2(\vrr)E^2(\vrr)d\vrr}.
\label{eq:gamma_bar_fca_int}
\end{align}
\noindent Substituting eq. (\ref{eq:TPA_bulk}) for $\gamma_{\text{TPA}}(\vrr)$, the modal loss rate due to free-carrier absorption in the porous silicon photonic crystals considered here can be written as
\begin{equation}
\label{eq:gamma_fca_bar_2}
\overline{\gamma}_{\text{FCA}} = \Gamma_{\text{FCA}}\left(\frac{\tau \sigma'_{\text{Si}} \beta'_{\text{Si}}}{2\hbar\omega_o} \frac{U^2}{V_{\text{FCA}}^2}\right),
\end{equation}
\noindent with effective confinement factor and mode volume defined as
\begin{align}
\Gamma_{\text{FCA}} &= \frac{\int_{\text{Si}} n^6(\vrr)E^6(\vrr) d\vrr}{\int n^6(\vrr)E^6(\vrr) d\vrr} \label{eq:Gamma_FC} \\
V_{\text{FCA}}^2 &=  \frac{ \left( \int{n^2(\vrr)E^2(\vrr) d\vrr}\right)^3}{\int{n^6(\vrr)E^6(\vrr) d\vrr}}. \label{eq:V_FCA}
\end{align} 

Equations (\ref{eq:gamma_fca_bar_2}) and  (\ref{eq:gamma_tpa_bar_2}) represent the total loss rate of photons from the cavity due to free-carrier and two-photon absorption, respectively.  These expressions depend only on material parameters, modal confinement factors $\Gamma_{\text{TPA,FCA}}$, effective mode volumes $V_{\text{TPA,FCA}}$, and the internal cavity energy $U$.  The modal parameters take account of the non-trivial geometry and field distribution of the cavity mode, and can be determined for a given mode from finite difference time domain simulations.  Including expressions (\ref{eq:gamma_fca_bar_2}) and  (\ref{eq:gamma_tpa_bar_2}) in $\gamma^i$,  eqs. (\ref{eq:U_NL} - \ref{eq:Q_i_P_NL}) can be solved iteratively for $Q_{i+P}$ and $K$, which characterize the on-resonance nonlinear response of the cavity for a given input power.

\subsection{Nonlinear and thermal dispersion}
\label{subsec:NL_thermal_disp_theory}
In addition to modifying the cavity quality factor, large cavity energy densities also modify the refractive index of the cavity, resulting in a power dependent resonance frequency. Here we consider the role of the Kerr effect, free-carrier dispersion, and heating due to linear and nonlinear absorption, on the dispersive response of the PC cavity.  
The refractive index shift induced through the processes considered here is a function of both space and internal cavity energy.  The renormalization of the resonant cavity frequency, resulting from small local perturbations in the refractive index, can be approximated using first order perturbation theory as 
\begin{equation}
\label{eq:d_omega}
\frac{\Delta \omega_o(U)}{\omega_o} = - \Delta \overline{n}(U),
\end{equation}
\noindent where the normalized modal index shift, $\Delta \overline{n}(U)$, is given by an average of the (normalized) local refractive index shift $\Delta n(\vrr)/n(\vrr)$, 
\begin{equation}
\label{eq:d_nbar}
\Delta \overline{n}(U) = \frac{\int \bigl(\frac{\Delta n(\vrr)}{n(\vrr)}\bigr) n^2(\vrr)E^2(\vrr) d\vrr}{\int n^2(\vrr)E^2(\vrr)d\vrr}.
\end{equation}
\noindent This energy dependent frequency shift, together with the energy dependent loss described in Section \ref{subsec:NL_abs_theory}, modifies the Lorentzian frequency dependence of the cavity response:
\begin{equation}
\label{eq:R_o_omega}
R_o(\omega) = 1 - \frac{4K(U)}{(1+K(U))^2}\frac{(\delta\omega/2)^2}{(\omega - \omega_o - \Delta\omega_o(U))^2 + (\delta\omega(U)/2)^2}.
\end{equation}
\noindent For a given input power $P_i$ and frequency $\omega$,  $U$ is given by
\begin{equation}
\label{eq:U_NL_omega}
U = \frac{P_{d}}{\gamma_{i+P}} = (1 - R_o(\omega))\frac{Q_{i+P}(U)}{\omega_{o}}  P_{i},
\end{equation}
\noindent where $P_{d} = (1 - R_o(\omega)) P_{i}$ is the frequency dependent dropped power in the resonant cavity.
 
For input powers sufficient to shift $\Delta \omega_o > \sqrt{3}\delta\omega/2$, the frequency response described by eq. (\ref{eq:R_o_omega}) is bistable, and can be exploited for applications including temperature locking and optical switching \cite{ref:Almeida1,ref:Almeida2,ref:Carmon}.  In order to solve eq. (\ref{eq:R_o_omega}) for the cavity response, it is necessary to derive expressions for each of the constituents of $\Delta \overline{n}$ as a function of $U$.  We begin with the Kerr effect.

\subsubsection{Kerr effect}
\label{subsubsec:KD_theory}
The time-averaged local index shift induced by the Kerr effect is
\begin{equation}
\label{eq:Kerr_shift_def}
\Delta n_{\text{Kerr}}(\vrr) =  n'_2(\vrr) \frac{1}{2}\epsilon_o n^2(\vrr)E^2(\vrr),
\end{equation}
\noindent where $n'_2(\vrr)$ is a material parameter, and is related to the usual $n_2$ coefficient relating intensity to refractive index shift \cite{ref:Boyd} by $n'_2 = (c/n_g) n_2$. In a silicon PC cavity the normalized modal index change due to the Kerr effect can be written as\footnote{This expression neglects the tensor nature of the third order susceptibility. A more general expression, which includes the tensor nature of the Kerr effect, is given in Ref.\ \cite{ref:Soljacic2}.}
\begin{equation}
\label{eq:dn_Kerr_bar}
\Delta \overline{n}_{\text{Kerr}}(U) = \frac{\Gamma_{\text{Kerr}}}{n_{\text{Si}}} \left( n'_{2,{\text{Si}}} \frac{U}{V_{\text{Kerr}}}\right),
\end{equation}
\noindent with $n'_{2,\text{Si}}$ and $n_{\text{Si}}$ the Kerr coefficient and linear refractive index of Si, respectively.  As both the Kerr effect and two photon absorption (TPA) share the same dependence on field strength, the confinement factor and effective mode volume associated with the Kerr effect are equal to those of TPA,
\begin{align}
\Gamma_{\text{Kerr}} &= \Gamma_{\text{TPA}}\\
V_{\text{Kerr}} &= V_{\text{TPA}}.
\end{align}
\subsubsection{Free-carrier dispersion}
\label{subsubsec:FCD_theory}
Dispersion due to free-carrier electron-hole pairs is given by\footnote{Experimental results\cite{ref:Soref} indicate that this Drude model must be modified slightly to accurately describe the contribution from hole free-carriers in silicon, which scales with $N_h^{0.8}$. For simplicity, we ignore this in the following analysis, and note that the modification is straightforward, and is included in later numerical results.}

\begin{equation}
\label{eq:dn_FCD}
\Delta n_{\text{FCD}}(\vrr) = - \zeta(\vrr) N(\vrr),
\end{equation}
\noindent where $\zeta(\vrr)$ is a material parameter with units of volume.  Following the derivation of $\overline{\gamma}_{\text{FCA}}$, the normalized modal index change is 
\begin{equation}
\label{eq:dn_fcd_bar}
\Delta \overline{n}_{\text{FCD}}(U) = -\frac{\Gamma_{\text{FCD}}}{n_{\text{Si}}}\left(\frac{\tau \zeta_{\text{Si}} \beta'_{\text{Si}}}{2\hbar\omega_o} \frac{U^2}{V_{\text{FCD}}^2}\right),
\end{equation}
\noindent with
\begin{align}
\Gamma_{\text{FCD}} &= \Gamma_{\text{FCA}}\\
V_{\text{FCD}} &= V_{\text{FCA}}.
\end{align}

\subsubsection{Thermal dispersion}
\label{subsubsec:thermal_theory}
It is also necessary to consider the effect of thermal heating due to optical absorption on the refractive index of the PC cavity.  The normalized modal index shift is given by
\begin{equation}
\label{eq:dn_thermal_th}
\Delta \overline{n}_{\text{th}} = \frac{\int{\Bigl(\frac{1}{n(\vrr)}\frac{dn}{dT}(\vrr) \Delta T(\vrr)\Bigr)n^2(\vrr)E^2(\vrr)d\vrr}}{\int{n^2(\vrr)E^2(\vrr) d\vrr}}.
\end{equation}
\noindent Here $\Delta T(\vrr)$ is the local temperature change due to the absorbed optical power density within the cavity,  and $dn/dT$ is a material dependent thermo-optical coefficient.  Neglecting differences in the spatial distributions\footnote{Heating due to each absorption process will result in a spatial temperature profile given by solving the heat equation.  In general the different processes have different spatial heating, and therefore temperature, profiles.} of the contributions to $\Delta T(\vrr)$ from the various absorption processes, and assuming that $\Delta T(\vrr)$ scales linearly with absorbed power density for a fixed spatial heating distribution, the modal index shift can be written as
\begin{align}
\label{eq:dn_th_bar}
\Delta \overline{n}_{\text{th}}(U) = \frac{\Gamma_{\text{th}}}{n_{\text{Si}}} \left(\frac{dn_{\text{Si}}}{dT}\frac{dT}{dP_{\text{abs}}}P_{\text{abs}}(U)\right)
\end{align}
\noindent where,
\begin{align}
\label{eq:P_abs}
P_{\text{abs}}(U) &= \left(\gamma_{\text{lin}} + \overline{\gamma}_{\text{TPA}}(U) + \overline{\gamma}_{\text{FCA}}(U^2)\right)U.
\end{align}
\noindent $\Gamma_{\text{th}} = \int_{\text{Si}} n^2E^2d\vrr / \int n^2E^2d\vrr$ is a confinement factor which accounts for the fact that only the semiconductor experiences an appreciable index shift, and $dT/dP_{\text{abs}}$ is the thermal resistance of the PC cavity which relates the mean modal temperature change to the total absorbed power.  In what follows we lump these two factors together, yielding an effective thermal resistance of the PC cavity.

From eqs. (\ref{eq:dn_Kerr_bar}), (\ref{eq:dn_fcd_bar}), and (\ref{eq:dn_th_bar}), the total modal index change and corresponding resonance frequency shift can be determined as a function of cavity energy.  The nonlinear lineshape described by eq. (\ref{eq:R_o_omega}) can then be calculated iteratively as a function of input power when combined with the power dependent loss model of sub-section \ref{subsec:NL_abs_theory}.  This is used below in Section \ref{sec:NL_meas} to estimate the scale of the different nonlinear processes in silicon PC microcavities. 

As a final comment we note that the above analysis has assumed a steady-state optical, carrier, and thermal distribution, whereas of significant practical interest for applications such high speed switching is the transient response of such structures.  Although the Kerr nonlinearity, two-photon absorption, free-carrier absorption, and free-carrier dispersion all depend on the electronic structure of the semiconductor material, the sub-micron geometry typical of photonic crystals can also play an important role.  For example, the surfaces introduced by the slab and air hole geometry of planar PC cavities can significantly modify the free-carrier lifetime $\tau$ compared to that in bulk material\cite{ref:Almeida1,ref:Claps,ref:Rong}.  Similarly, the thermal response time scales inversely with the spatial scale of the optically absorbing region, and depends upon the geometry and material dependent thermal properties of a given structure \cite{ref:Boyd}.  Although not the focus of the work presented here, an inkling of these effects is seen in the sub-nanosecond estimated effective free-carrier lifetime in the silicon PC cavity studied below. 

\section{Efficient coupling into a high-Q cavity mode}
\label{sec:Lin_meas}

\begin{figure}[ht]
\begin{center}
\epsfig{figure=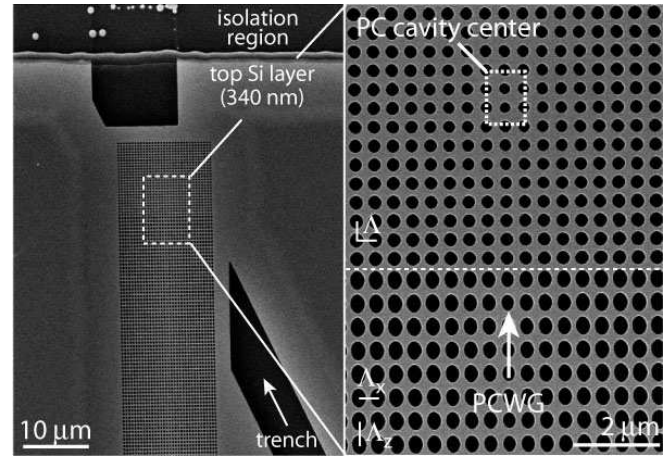, width=1.0\linewidth}
\caption{SEM image of an integrated PCWG-PC cavity sample.  The PC cavity and PCWG have lattice constants $\Lambda \sim 430$ $\text{nm}$, $\Lambda_{x} \sim 430$ $\text{nm}$, and $\Lambda_{z} \sim 550$ $\text{nm}$.  The surrounding silicon material has been removed to form a diagonal trench and isolated mesa structure to enable fiber taper probing.}
\label{fig:SEM}
\end{center}
\end{figure}

The integrated PC cavity-PCWG devices described in Section \ref{sec:couple_theory_linear} were fabricated in an optically thin layer (thickness 340 nm) of silicon as described in Ref. \cite{ref:Barclay4}.  A typical device is shown in Fig. \ref{fig:SEM}, which also shows the regions in which unpatterned silicon was removed to allow taper probing of the PC devices.  In addition to isolating the PC devices on a mesa of height $\sim10$ $\mu\text{m}$, a trench extending diagonally from the cavity was defined.  This allowed the cavity to be probed directly by the fiber taper as in Ref.\ \cite{ref:Srinivasan4}.  Indirect, yet much more efficient coupling to the PC cavity, was performed using the scheme described in Sec.\ \ref{sec:couple_theory_linear} by aligning the fiber taper along the axis of the PCWG and coupling through the PCWG into the PC cavity.  The fiber taper was mounted in a ``u''-shape on a DC motor $z$-stage with $50$ nm encoder resolution, providing accurate vertical placement of the taper above the Si chip.  In the measurements described below the taper was held fixed in the $x$-$y$ plane parallel to the Si chip surface, and in-plane positioning of the Si chip was performed using a pair of DC motor stages with similar encoder resolution.  A rotation and goniometer stage, mounted to the $x$-$y$ stages, was used to align the in-plane and vertical angle of the sample relative to the taper, respectively.  The entire sample and taper system was enclosed in an acrylic box to reduce taper fluctuations due to air currents in the room.  Further details of the fiber taper fabrication and mounting geometry can be found in Refs. \cite{ref:Srinivasan4,ref:Barclay5,ref:Barclay4}. 

\begin{figure}[ht]
\begin{center}
\epsfig{figure=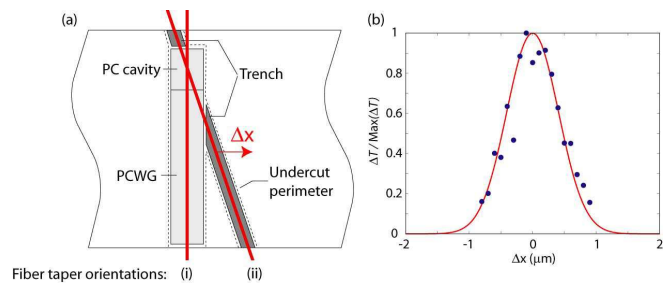, width=1.0\linewidth}
\caption{(a) Illustration of the device and fiber taper orientation for (i) efficient PCWG mediated taper probing of the cavity, and (ii) direct taper probing of the cavity.  (b) Normalized depth of the transmission resonance ($\Delta\overline{T}$) at $\lambda_o \sim 1589.7$, as a function of lateral taper displacement relative to the center of the PC cavity, during direct taper probing (taper in orientation (ii)).}\label{fig:cav_probe}
\end{center}
\end{figure} 

A fiber-coupled swept wavelength (1565 - 1625 nm) laser source was used to measure the wavelength dependent forward transmission ($\overline{T}$) through the fiber taper, and a fiber splitter was used to monitor the signal in the backward propagating fiber taper mode ($\overline{R}$).  With the taper aligned with the etched trench, the spectral and spatial properties of the PC cavity modes were probed directly (Fig.\ \ref{fig:cav_probe}(a), taper position (ii)), as in Ref. \cite{ref:Srinivasan4}.  The trench prevents the fiber taper from interacting with the unpatterned silicon, and light is coupled directly from the fiber taper into the high-Q PC cavity modes.  Although this coupling is inefficient ($\Delta\overline{T}=$1-10$\%$, $I \ll 1$), it allowed the frequency of the $A_{2}^0$ cavity mode to be independently determined. 

In the device studied here, when the taper was aligned with the trench and positioned $\sim 500$ nm above the cavity, a sharp dip in $\overline{T}$ was observed at a wavelength of $\lambda_o \sim 1589.7$ nm. It was confirmed that this was due to coupling to a localized cavity mode by studying the depth of the resonance as the taper was displaced laterally ($|\Delta x| > 0$) relative to the center of the PC-cavity.  The measured normalized resonance depth as a function of taper displacement is shown in Figure \ref{fig:cav_probe}(b), and has a halfwidth of 480 nm, consistent with previous studies of the localized $A_{2}^0$ cavity mode\cite{ref:Srinivasan4}.

The fiber taper was then aligned above and parallel to the PCWG (Fig.\ \ref{fig:cav_probe}(a), taper position (i)).  At taper-PCWG phase-matching wavelengths, $\overline{T}$ decreases resonantly as power is coupled from the taper into the PCWG; coupling to the $\text{TE}_{1}$ PCWG mode was verified by studying the dispersive and spatial properties of the coupling, as in Ref. \cite{ref:Barclay4}.  The fiber taper-PCWG coupling bandwidth was adjusted to overlap with the wavelength of the $A_{2}^0$ cavity mode using two mechanisms.  Coarse tuning was obtained by adjusting, from sample to sample, the nominal hole size and longitudinal lattice constant ($\Lambda_{z}$) of the PCWG.  Fine tuning of the coupler's center wavelength over a 100 nm wavelength range was obtained by adjusting the position, and hence diameter, of the fiber taper region coupled to the PCWG \cite{ref:Barclay4}.  Different degrees of cavity loading were also studied by adjusting the number of periods (9-11) of air holes between the center of the PC cavity and the end of the PCWG.  In the device studied below (shown in Fig. \ref{fig:SEM}) the PC cavity was fabricated with 9 periods on the side adjacent the PCWG and 18 periods on the side opposite the PCWG . 

\begin{figure}[t]
\begin{center}
\epsfig{figure=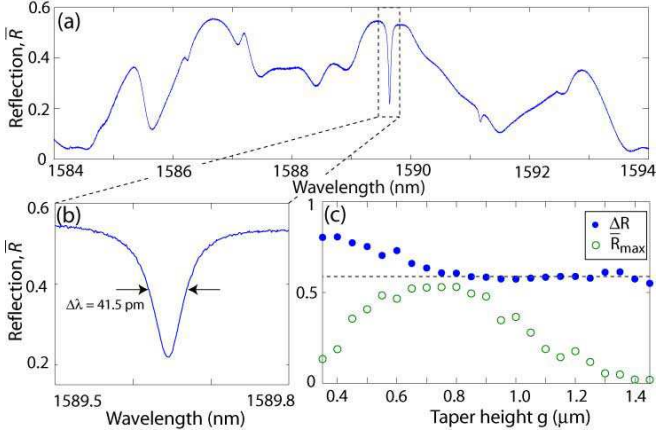, width=1.0\linewidth}
\caption{ (a) Measured reflected taper signal as a function of input wavelength (taper diameter $d \sim 1$ $\mu$m, taper height $g = 0.80$ $\mu$m).  The sharp dip at $\lambda \sim 1589.7$ nm, highlighted in panel (b), corresponds to coupling to the $A_{2}^0$ cavity mode. (c) Maximum reflected signal (slightly detuned from the $A_{2}^0$ resonance line), and resonance reflection contrast as a function of taper height.  The dashed line at $\Delta R = 0.6$ shows the PCWG-cavity drop efficiency, which is independent of the fiber taper position for $g \ge 0.8$ $\mu$m.}\label{fig:data}
\end{center}
\end{figure} 

Figure \ref{fig:data}(a) shows the normalized reflected fiber signal, $\overline{R}$, for a taper diameter $d\sim1$ $\mu\text{m}$, which aligns the taper-PCWG coupler bandwidth with the $A_{2}^0$ PC cavity mode wavelength.  This signal is normalized to the taper transmission in absence of the PCWG, and since light passes through the taper-PCWG coupler twice, is given by $\overline{R} = \eta_{wg}^2 R_{o}$, where $\eta_{wg}$ is the taper-PCWG coupling efficiency.  Note that both $R_{o}$ and $\eta_{wg}$ are frequency dependent.  In Figure \ref{fig:data}(a), the peak in $\overline{R}$ around $\lambda \sim 1590$ nm corresponds to the phase-matched point of the fiber taper and the $\text{TE}_{1}$ PCWG mode.  From the peak value of $\overline{R}_{\text{max}} = 0.53$, a maximum taper-PCWG coupling efficiency of $\eta_{wg} \sim 73 \%$ is estimated, where the off-resonant $R_{o}$ is taken to be unity.  This value is lower than the $97\%$ obtained in previous work \cite{ref:Barclay5} due to coupling to additional higher-order (normal to the Si slab) PCWG modes which interfere with the coupling to the fundamental $\text{TE}_{1}$ PCWG mode for strong taper-PCWG coupling.  This can be avoided in future devices by increasing the nominal PCWG hole size relative to that in the PC-cavity or reducing the Si slab thickness, effectively freezing out the higher-order PCWG modes \cite{ref:Barclay4}.

The sharp dip in reflection at $\lambda \sim 1589.7$ nm, shown in detail in Fig. \ref{fig:data}(b), corresponds to resonant excitation of the $A_{2}^0$ PC cavity mode, as confirmed by the direct fiber probing of the cavity described above. The other broad features in $\overline{R}$ correspond to weak Fabry-P\'{e}rot effects of the PCWG, and weak interference between the TE-1 mode and higher order PCWG modes.  The reflected fiber taper signal as a function of taper-PCWG gap height, $g$, is shown in Fig. \ref{fig:data}(c). For $g \ge 0.8$ $\mu\text{m}$, $\overline{R}_{\text{max}}$ increases with decreasing $g$ as the coupling from the fiber taper to the $\text{TE}_{1}$ PCWG mode becomes stronger.  The reflection contrast, $\Delta R = 1-R_{o}(\omega_o) = (\overline{R}_{\text{max}} - \overline{R}(\omega_o))/\overline{R}_{\text{max}}$, remains constant, since the PCWG-cavity interaction is independent of the fiber taper to PCWG coupling.  For smaller taper-PCWG gap heights, $g < 0.8$ $\mu$m, fiber taper coupling into higher order PCWG modes and radiation modes becomes appreciable, and  $\overline{R}_{\text{max}}$ decreases for decreased taper height.  The corresponding increase in $\Delta R$ seen in Fig. \ref{fig:data}(c) is a result of interference between the $\text{TE}_{1}$ mode and higher-order PCWG modes which are excited and collected by the taper, and is not a manifestation of improved coupling between the $\text{TE}_{1}$ PCWG mode and the $A_{2}^0$ PC cavity mode.  Direct coupling between the taper and the cavity is negligible here.

From a Lorentzian fit to the $A_{2}^0$ cavity resonance dip in $R_{o}(\omega)$, the normalized on-resonance reflected power is estimated to be $R_o(\omega_{o}) = 0.40$, corresponding to an undercoupled $K=0.225$.  The loaded quality factor as measured by the reflected signal linewidth is $Q_{T} = 3.8 \times 10^4$.  Substituting these values into eq. (\ref{eq:Q_i+P}) gives the cavity mode quality factor due to parasitic loading and intrinsic losses, $Q_{i+P} = 4.7 \times 10^4$.  Previous measurements of similar PC cavity devices without an external PCWG load yielded intrinsic quality factors of $4 \times 10^4$ \cite{ref:Srinivasan4}, strongly indicating that the parasitic loading of the PC cavity by the PCWG is minimal, and $I \sim 1$ for this PC cavity-waveguide system.  The high ideality of this coupling scheme should be contrasted with previous direct taper measurements of the PC cavities \cite{ref:Srinivasan4}, whose coupling was limited to a maximum value $K = 0.018$, with an ideality of $I\sim0.035$ (corresponding to a resonance depth of $7\%$, $Q_{T} = 2.2 \times 10^4$).

The efficiency of power transfer from the fiber taper into the PC cavity is given by $\eta_{in} = \eta_{wg}\Delta R \approx 44 \%$.  This corresponds to the \emph{total} percentage of photons input to the fiber taper which are dropped by the PC cavity.  Based upon these measurements, in the case of an internal cavity source such as an atom or a quantum dot, the efficiency of light collection into the fiber taper for this PC cavity system would be $\eta_{out} = \eta_{wg}\eta_{0} \approx 13\%$ ($\eta_{0} \approx 18\%$).  As the tapers themselves are of comparably very low loss, with typical losses associated with the tapering process less than $10\%$, these values accurately estimate the \emph{overall} optical fiber coupling efficiency.   Finally, note that previous measurements of near-ideal coupling between the fiber taper and PCWG \cite{ref:Barclay5} indicate that by adjusting the PCWG as described above to improve $\eta_{wg}$, $\eta_{in}$ and $\eta_{out}$ can be increased to $58\%$ and $18\%$, respectively.  More substantially, adjustments in the coupling parameter $K$ towards over-coupling by decreasing the number of air-hole periods between the PC cavity and the PCWG can result in significant increases in $\eta_{in}$ and $\eta_{out}$ with minimal penalty in loaded $Q$-factor for $I \sim 1$.

\section{Nonlinear measurements}
\label{sec:NL_meas}
\begin{figure}[ht]
\begin{center}
\epsfig{figure=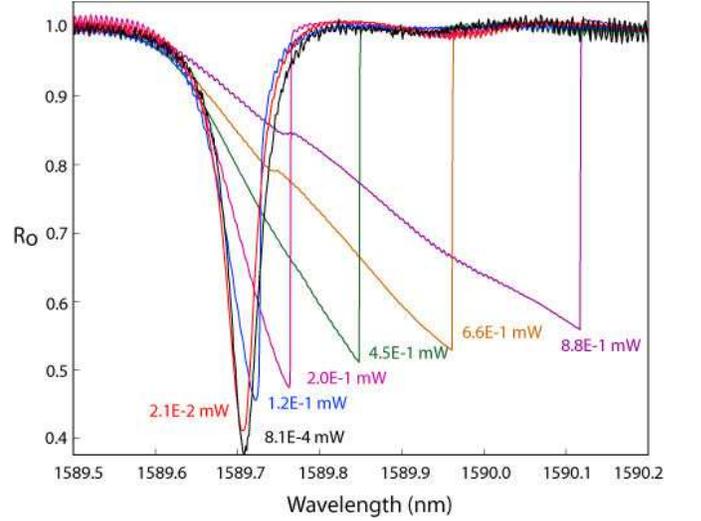, width=1.0\linewidth}
\caption{ (a) Measured cavity response as a function of input wavelength, for varying PCWG power  (taper diameter $d \sim 1$ $\mu$m, taper height $g = 0.80$ $\mu$m).}\label{fig:data_scan_NL}
\end{center}
\end{figure}

The nonlinear response of the PC cavity was studied by measuring the dependence of the reflected signal lineshape on the power input to the PCWG.  Figure  \ref{fig:data_scan_NL} shows wavelength scans of the cavity response $R_o$ for varying power, $P_t$, input to the fiber taper.  Each scan was obtained by dividing the normalized reflected signal, $\overline{R}$, by the slowly varying taper-PCWG coupler lineshape, $\eta_{wg}^2(\omega)$.  In all of the measurements, the fiber taper was aligned near the optimal taper-PCWG coupling position, and the wavelength of the laser source was scanned in the direction of increasing $\lambda$.   $P_t$ was determined by taking taper insertion loss into account, and measuring the taper input power with a calibrated power meter.  

Increasing the power in the fiber taper, and consequently the PCWG, results in three readily observable changes in $R_o$: (i) a decrease in the resonance contrast, $\Delta R_o$, (ii) a shift $\Delta\omega_o$ in the resonance frequency $\omega_o$, and (iii) broadening and asymmetric distortion of the resonance lineshape, eventually leading to a ``snap'' in the reflection response characteristic of bistability \cite{ref:Gibbs}.  Here we use the theory presented in Section \ref{sec:NL_theory} to show that these features are due to nonlinear absorption and dispersion in the PC cavity.

\begin{figure}[ht]
\begin{center}
\epsfig{figure=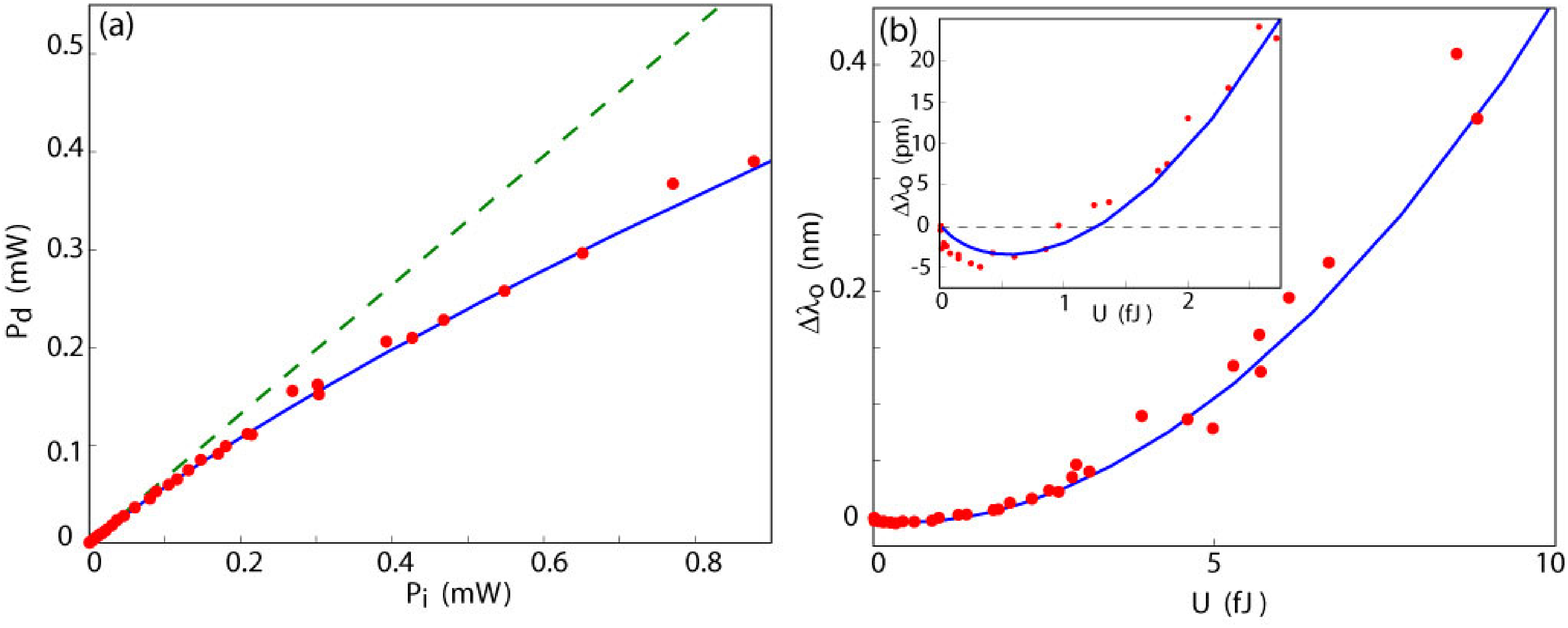, width=1.0\linewidth}
\caption{ (a) Power dropped ($P_{d}$) into the cavity as a function of power in the PCWG ($P_{i}$). The dashed line shows the expected result in absence of nonlinear cavity loss. (b) Resonance wavelength shift as a function of internal cavity energy.  Solid blue lines in both figures show simulated results.}\label{fig:data_Pd_dlambda_0}
\end{center}
\end{figure} 

Figure \ref{fig:data_Pd_dlambda_0}(a) shows $P_d$, the on-resonance power dropped into the PC cavity, as a function of $P_i$, the power incident on the cavity from the PCWG.  $P_d$ is measured from $P_d = \Delta R_o(P_i) P_i$, and $P_i$ is related to the taper input power by $P_i = \eta_{wg}(\omega_o)P_{t}$.  For small $P_i$, $P_d$ increases with a constant slope equal to the ``cold cavity'' value of $\Delta R_o = 0.60$ measured in Section \ref{sec:Lin_meas}.  For larger $P_i$, $P_d$ becomes sub-linear versus $P_{i}$ as loss due to nonlinear absorption becomes appreciable compared to the other loss channels of the PC cavity.  In the context of the analysis of Section \ref{sec:NL_theory}, $\gamma^i$ increases with increasing $P_d$, degrading $K$, and decreasing $\Delta R_o$ (for $K < 1$). From the ``cold cavity'' $\eta_0$ and $Q_T$ measured in the previous section, the power dependent $Q_{i+P}(P_i)$ can be extracted from $\Delta R_o(P_i)$ through the relation:
\begin{equation}
\label{eq:Q_i+P_P_i}
Q_{i+P}(P_i) = K(\Delta R_o(P_i))\frac{Q_T(P_i=0)}{\eta_0(P_i= 0)}.
\end{equation}
\noindent Equation (\ref{eq:Q_i+P_P_i}) is useful for powers where nonlinear effects distort the Lorentzian lineshape, and  $\lambda_o/\delta\lambda$ is not an accurate measure of $Q_T(P_i)$.  Using eqs. (\ref{eq:Q_i+P_P_i}) and (\ref{eq:U}), the internal cavity energy, $U$, can be calculated from $P_i$ and $\Delta R_o$.

Figure \ref{fig:data_Pd_dlambda_0}(b) shows a plot of the measured\footnote{Note that the sharp transition edge associated with optical bistability occurs at the cavity resonance wavelength when scanning from blue to red, thus an accurate measure of $\Delta \lambda_o$ can be made, even if one of $\delta\lambda$ cannot.} $\Delta \lambda_o$, the resonance wavelength shift, as a function of $U$.  This plot has several noteworthy properties.  First, the wavelength shift is nonlinear in $U$, indicating that nonlinear processes such as free-carrier dispersion and heating through nonlinear absorption must be taking effect.  Also, for small $U$ the resonance wavelength is seen to blue shift.  In the $1550$ $\text{nm}$ wavelength band of operation both $dn_{\text{Si}}/dT$ and $n_{2,\text{Si}}$ are $>0$, while $d(\Delta n_{\text{FCD}})/dU < 0$, indicating that free-carrier dispersion is the dominant dispersive process at low input powers.  For $U > 0.34$ $\text{fJ}$ ($P_d > 10$ $\mu\text{W}$), the resonance wavelength begins to red shift, indicating that thermal or Kerr effects dominate for large internal cavity energy.  Also, note that for a stored cavity energy as low as $U \sim 3$ fJ ($P_d \sim 100$ $\mu$W) the cavity response is bistable with $\Delta\lambda_o = 35$ $\text{pm} \sim \sqrt{3} \delta\lambda/2$.  

In order to estimate the contributions of the various nonlinear processes to the effects discussed above, the absorptive, $P_d(P_i)$, and dispersive, $\Delta\lambda(P_i)$, data were fit using the model presented in Section \ref{sec:NL_theory}.  Specifically, eqs. (\ref{eq:U_NL}-\ref{eq:Q_i_P_NL}) were solved for $P_d$ and $U$ as a function of $P_i$, and eq. (\ref{eq:d_omega}) was used to calculate $\Delta\lambda_o$.  The free parameters in this model were taken as: (i) the effective free-carrier lifetime, $\tau$, (ii) the effective thermal resistance of the PC cavity, $\Gamma_{\text{th}}dT/dP_{\text{abs}}$, and (iii) the fraction of the ``cold cavity'' loss which is due to linear absorption (as opposed to radiation), $\eta_{\text{lin}} = \gamma_{\text{lin}}/(\gamma_{\text{lin}} + \gamma_{\text{rad}})$.   The material and modal constants used are listed in Table \ref{tab:parameters}. 

\begin{figure}[t]
\begin{center}
\epsfig{figure=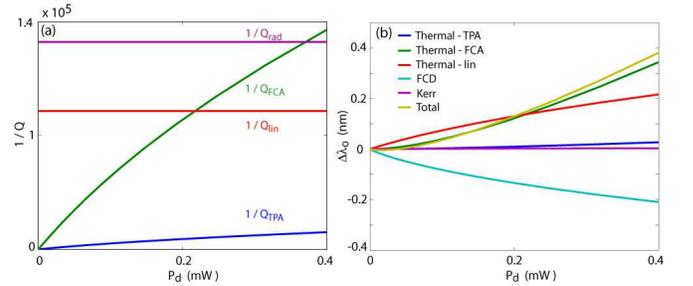, width=1.0\linewidth}
\caption{(a) Simulated effective quality factors for the different PC cavity loss channels as a function of power dropped into the cavity.  (b) Contributions from the modeled dispersive processes to the PC cavity resonance wavelength shift as a function of power dropped into the cavity.  (Simulation parameters: $\eta_{\text{lin}} \sim 0.40$, $\Gamma_{\text{th}}dT/dP_{\text{abs}} = 27$ K/mW, $\tau^{-1} \sim 0.0067 + (1.4\times 10^{-7}) N^{0.94}$ where $N$ has units of $\text{cm}^{-3}$ and $\tau$ has units of ns).
}\label{fig:model_Q_dlambda_0}
\end{center}
\end{figure}

As has been observed in studies of silicon optical waveguides\cite{ref:Liang}, we find that a strong dependence of $\tau$ on carrier density is required for our model to accurately reproduce \emph{both} the dispersive and absorptive data represented in Figures \ref{fig:model_Q_dlambda_0}(a) and (b).  In order to account for a carrier density dependent lifetime in our model the following procedure was used.  With $\Gamma_{\text{th}}dT/dP_{\text{abs}}$ and  $\eta_{\text{lin}}$ held fixed, $\tau(P_{i})$ was determined for each input power from a least squares fit to $\Delta\lambda_o(P_i)$ and $P_d(P_i)$\footnote{Note that since we have two data points for each input power, one dispersive and one absorptive, there will be an optimum $\tau(P_{i})$ with non-zero residual error.}.  This procedure was repeated for a range of values for $\Gamma_{\text{th}}dT/dP_{\text{abs}}$ and  $\eta_{\text{lin}}$.  For a fixed value of $\eta_{\text{lin}}$, the fits were robust in $\Gamma_{\text{th}}dT/dP_{\text{abs}}$ with the sum of the least square residual of $\tau(P_{i})$ clearly minimized for an optimal value of $\Gamma_{\text{th}}dT/dP_{\text{abs}}$.  This procedure, however, was only found to constrain $\eta_{\text{lin}} > 0.15$.  Within this range of $\eta_{\text{lin}}$ the quality of the fits does not change significantly, with the optimal functional form of $\tau$ changing slightly and the optimal value of $\Gamma_{\text{th}}dT/dP_{\text{abs}}$ varying between $\sim 15-35$ K/mW.  Based on estimates of $\eta_{\text{lin}}$ from studies of loss in silicon microdisk resonators fabricated using the same SOI wafers and the same processing techniques\cite{ref:Borselli}, and by comparing the etched surface area seen by the PC cavity mode to that seen by a microdisk mode, we chose to use $\eta_{\text{lin}} \sim 0.40$ for the PC cavity.  With this value of $\eta_{\text{lin}}$ the optimal value of the effective cavity thermal resistance, $\Gamma_{\text{th}}dT/dP_{\text{abs}}$, was found to be $27$ K/mW, of the same order of magnitude as the result calculated in Ref. \cite{ref:OBrien3} for a similar membrane structure.  Finally, the point-by-point least-squared optimum values of $\tau(P_{i})$ were then fit with a smooth curve of functional form $\tau^{-1} = A + BN^{\alpha}$ as a function of the effective free-carrier density $N$. Using this $\tau(N)$ fit, smooth fits to measured $P_d(P_i)$ and $\Delta\lambda(P_i)$ were obtained, shown as solid blue lines in Fig. \ref{fig:data_Pd_dlambda_0}.

\renewcommand{\arraystretch}{1.175}
\begin{center}
\begin{table}
\caption{Fixed parameters used in the model.}
\label{tab:parameters}
\begin{tabular}{ l  c  c  c }
  \hline\hline 
  Parameter             & Value                         & Units                                 & Source                        \\ \hline\hline
  $V_{\text{TPA}}$      & $4.90$                        & $(\lambda_o / n_{\text{Si}})^3$       & FDTD(b)                 	\\ \hline
  $V_{\text{FCA}}$      & $3.56$                        & $(\lambda_o / n_{\text{Si}})^3$       & FDTD(b)               	\\ \hline
  $\Gamma_{\text{TPA}}$ & $0.982$                       & -                                     & FDTD(b)       	        \\ \hline
  $\Gamma_{\text{FCA}}$ & $0.997$                       & -                                     & FDTD(b) 	                \\ \hline
  $n_{\text{Si}}$       & $3.45$                        & -                                     & \cite{ref:Cutolo,ref:Soref}   \\ \hline
  $\sigma_{\text{Si}}$  & $14.5\times10^{-22}$          & $\text{m}^{2}$                        & \cite{ref:Cutolo,ref:Soref}   \\ \hline
  $\zeta_{\text{Si}}^e$ & $8.8\times10^{-28}$           & $\text{m}^{3}$                        & \cite{ref:Cutolo,ref:Soref}   \\ \hline
  $\zeta_{\text{Si}}^h$ & $4.6\times10^{-28}$           & $\text{m}^{3}$                        & \cite{ref:Cutolo,ref:Soref}   \\ \hline
  $n_{2,\text{Si}}$     & $4.4\times10^{-18}$           & $\text{m}^2 \cdot \text{W}^{-1}$      & \cite{ref:Dinu}               \\ \hline
  $\beta_{\text{Si}}$   & $8.4\times10^{-12}$           & $\text{m} \cdot \text{W}^{-1}$        & \cite{ref:Dinu} (a)          \\ \hline
  $dn_{\text{Si}}/dT$   & $1.86\times10^{-4}$           & $\text{K}^{-1}$                       & \cite{ref:Cocorullo}          \\ \hline\hline
\end{tabular}
\begin{tablenotes}
\item (a) Average of the two quoted values for Si$\langle 110\rangle$ and Si$\langle 111 \rangle$. \\
\item (b) Calculated from FDTD generated fields of the $A_2^0$ cavity mode of the graded square lattice cavity studied here. \\
\end{tablenotes}
\end{table}
\end{center}
\renewcommand{\arraystretch}{1.0}

\begin{figure}[ht]
\begin{center}
\epsfig{figure=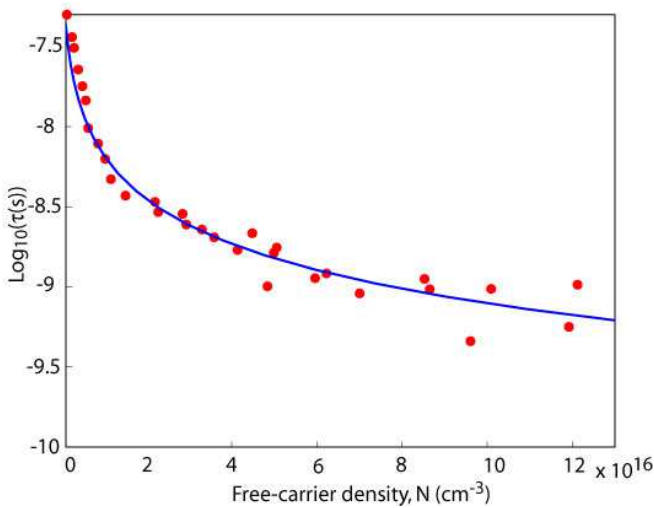, width=1.0\linewidth}
\caption{Dependance of free-carrier lifetime on free-carrier density (red dots) as found by fitting $\Delta\lambda_o(P_i)$ and $P_d(P_i)$ with the constant material and modal parameter values of Table \ref{tab:parameters}, and for effective PC cavity thermal resistance of $\Gamma_{\text{th}}dT/dP_{\text{abs}} = 27$ K/mW and linear absorption fraction $\eta_{\text{lin}}=0.40$.  The solid blue line corresponds to a smooth curve fit to the point-by-point least-squared fit data given by $\tau^{-1} \sim 0.0067 + (1.4\times 10^{-7}) N^{0.94}$, where $N$ is in units of cm$^{-3}$ and $\tau$ is in ns.}\label{fig:tau_FC}
\end{center}
\end{figure}

The various components of the total cavity loss rate and resonance shift based upon the above fits to the measured data are shown in Fig. \ref{fig:model_Q_dlambda_0}.  It can be seen that although TPA does not dominate the PC cavity response, the free-carriers it generates and the resulting free carrier dispersion and absorption drive the nonlinear behavior of the silicon PC cavity at low and high input powers, respectively. The fit effective free-carrier lifetime, shown in Fig. \ref{fig:tau_FC}, shows similar characteristics to that obtained by Tsang, et al. \cite{ref:Liang}, demonstrating a significant fall-off in $\tau$ for large $N$, but with a smaller saturated lifetime. Both the pronounced decay in $\tau$ and the low $\sim 0.5$ ns value of the high-carrier density free-carrier lifetime are significantly different from that found in bulk Si, and are most likely related to carrier diffusion and surface effects owing to the extremely large surface-to-volume ratio of the PC cavity, the small length scales involved ($\sim 200$ nm feature size), and the small size scale of the optical mode\cite{ref:Claps}.  This small effective free-carrier lifetime is consistent with other recent experimental results of highly porous silicon optical structures\cite{ref:Almeida1,ref:Rong}.  It should, however, be noted that the bulk Si TPA coefficient was used in modelling the nonlinear response of the PC cavity, which given the above comments may not be accurate due to surface modification of TPA.  As the effects of free-carrier lifetime and two-photon absorption on the behaviour of the dispersive and absorptive nonlinear response of the PC cavity are somewhat intertwined, further studies will be necessary to concretely separate these two phenomena in porous Si structures such as the photonic crystals of this work. 

\section{Summary}
\label{sec:Summary}
In conclusion, we have demonstrated an optical fiber coupling scheme to efficiently source and collect light from high-Q ultra-small mode volume PC cavities.  By employing a PCWG which supports a mode which is simultaneously spatially mode matched with the high-Q PC cavity mode of interest and phase matched with an optical fiber taper, efficient fiber-PCWG-cavity coupling is enabled.  A total fiber-to-cavity coupling efficiency of $44\%$ is demonstrated, which for the case of an internal cavity emitter corresponds to a radiated photon collection efficiency of $13\%$.  These values are not fundamental limits of this technique, and can be improved through fine tuning of the photonic crystal.  The efficiency of this fiber coupling method was then exploited to probe the steady-state nonlinear optical properties of the PC cavity.  The effect of two-photon absorption, free-carrier absorption and dispersion, Kerr self-phase modulation, and thermo-optic dispersion, on the response of the PC cavity was considered.  Optical bistability at fiber input powers of $250$ $\mu$W was observed, and a free-carrier lifetime with high carrier density value as low as $\sim 0.5$ ns is inferred from nonlinear absorptive and dispersive measurements of the PC cavity.  Along with applications to nonlinear optics, this optical fiber evanescent coupling based approach should be useful for future experiments in integrated micro-optics with photonic crystals, and in particular to cQED systems employing photonic crystals, where quantum computing and communication protocols demand high optical fidelity.

The authors thank Matthew Borselli for helpful discussion.  This work was supported by the Charles Lee Powell foundation.  KS would also like to thank the Hertz foundation for its support through a graduate fellowship.

\bibliography{/home/paul/biblio/PBG.bib}

\end{document}